\documentstyle[11pt,newpasp,twoside]{article}

\input epsf 

\newcommand{\oversim}[2]{\protect{\mbox{\lower0.5ex\vbox{%
   \baselineskip=0pt\lineskip=0.2ex
   \ialign{$\mathsurround=0pt #1\hfil##\hfil$\crcr#2\crcr\sim\crcr}}}}} 
\newcommand{\simless} {\mbox{$\,\mathrel{\mathpalette\oversim<}\,$}} 
\newcommand{\tcr}{t_{\rm cr}}

\newcommand{\erf}{{\rm erf}}

\def\edcomment#1{\iffalse\marginpar{\raggedright\sl#1\/}\else\relax\fi}
\reversemarginpar

\begin{document}
\title{Impact of Gas Removal on the Evolution of Embedded Clusters} 
 \author{Christian Boily}
\affil{Astronomisches Rechen-Institut, Heidelberg}
\author{Pavel Kroupa} 
\affil{Institut f\"ur Theoretische Physik und Astrophysik, Kiel} 

\begin{abstract}
We study both analytically and numerically the disruptive effect of
instantaneous gas removal from an embedded cluster. We setup a
calculation based on the stellar velocity distribution function, to
compute the fraction of stars that remain bound once the cluster has
ejected the gas and is out of equilibrium. We find tracks of bound
mass-fraction vs star formation efficiency similar to those obtained
with N-body calculations. We use these to argue that embedded clusters
must develop high-binding energy cores if they are to survive as bound
clusters despite a star formation rate as low as 20\% or lower
suggested by observations.
\end{abstract}

\section{Introduction}
Most if not all stars are formed in embedded clusters and associations, 
 and therefore this is likely the predominant 
mode of star formation contributing to the Galactic-field population.
To understand  how such aggregates form and evolve remains a
severe challenge however, since no  theory exists yet 
that accounts in detail for the (presumably) simpler case of the formation of
individual stars. On the observational front, surveys of star-forming
regions suggest that the mass-fraction of gas used up at birth does
not exceed 10 to 20\% of the total (Lada 1999).  This low
star-formation efficiency (sfe) implies that young stars or clusters
of stars should be embedded in gas. Yet populous clusters with ages $\simless 1$~Myr
(e.g. the Orion Nebula Cluster, R136 in 30~Doradus) are already void
of gas.  The fraction of gas left behind after the epoch of star
formation must therefore be removed {\it quickly} (well before the
first supernova explodes) to reveal a bare cluster. By the time this
occurs, most of the cluster must be assembled. Hydrodynamic collapse solutions 
point to  the rapid formation of stellar cores followed by time-dependent 
accretion (Boily \& Lynden-Bell 1995). Stellar masses are then accrued over 
a period $10^5 - 10^6$ years, long before the gas-evacuation timescale. 
Therefore gas-evacuation itself inhibits further star formation in the 
cluster. The gas is driven out by  OB stars that
culminate the star-formation process.  Stellar winds from these OB
stars yield a momentum flux ${\cal F} \sim 8\times 10^{-3} M_{\sun}\,
{\rm km\, sec}^{-1} {\rm yr}^{-1}$ (Churchwell 1999). This is
sufficient to blow out gas from an $10^3\,M_\odot$, 1 pc radius
embedded cluster in $\approx 10^5$ years, which is comparable to its
crossing time $\tcr = 2 R/\sigma$.  
In addition, the ionising radiation heats the
gas to $10^4$~K, causing an overpressure and expansion at the
sound velocity $\approx 10$~km/s.  
  We want to establish how significant mass-loss 
 affects the structure of a cluster, in particular what
{\it fraction} of the stars remain bound. 

This contribution marks the start of a research programme aiming at 
 identifying and quantifying the fundamental physical processes responsible
for the formation of open clusters. This
programme is inspired by the finding from high-precision $N$-body
computations that open clusters can form readily despite low
star-formation efficiencies,  in contradiction to previous results
(Kroupa, Aarseth \& Hurley 2001). Three hypothesis are raised that may
explain these $N$-body results. (i) Either two-body encounters {\it
during} the expansion phase of the cluster after gas expulsion
condense a part of the radial flow to a bound entity, or (ii)
encounters between the ubiquitous binary systems {\it during} the
radial outflow affect the condensation, or (iii) two-body and
binary--binary encounters {\it before} gas-expulsion (i.e. during the
embedded phase) cause the segregation of a tightly bound core that
resists expansion when the gas leaves the system. 

In order to weigh the relative contributions from each of points (i)-(iii), 
 all drawn from particle-particle (collisional) evolution, 
we seek first  to isolate the salient features of   collisionless, smooth 
 open cluster evolution. We 
take an analytic approach to the problem, based on the velocity distribution 
function of stars. 
By applying a new, fast and self-consistent iterative method for 
computing the 
fraction of bound stars, we show that to account
for observations of low sfe, clusters must develop strongly bound
cores to avoid dissolving completely owing to mass loss. 
  In addition to our analytical approach we use numerical
calculations to assess its range of applicability.

\section{The problem} 
A cluster of stars forms converting a fraction $\epsilon$ of the gas
into stars in the process. Stellar winds or a supernova event blow/s
out the remainder on a timescale $\tau \ll \tcr$. What fraction of the
stars remain to form a bound cluster ? How is the equilibrium profile
linked to the initial mass density ? Since the star formation epoch
will have lasted as long as or longer than a cluster dynamical time
$\tcr$, the system as a whole is close to virial equilibrium
before gas is expelled.

Hills (1980) argued that equilibrium, self-gravitating stellar
systems would expand or even dissolve if half the mass is lost very
quickly: stars then preserve their kinetic energy established under a
deeper potential well, hence may escape if their binding energy
becomes positive. We write the total energy

\[ E = - \kappa \frac{GM^2}{R} + \frac{1}{2} M\langle v^2\rangle < 0\ ,\]
where $M, R$ are the mass and radius of a uniform-density spherical
distribution of mean square velocity $\langle v^2\rangle$, and
$\kappa = 3/5$ for  the particular geometry and density profile
considered.  Before gas-expulsion the total mass in gas and stars is
$M  = M_{\rm init} = 
M_{\rm gas}+M_\star$, while after the gas is expelled
$M=M_\star$ but the velocity dispersion, $\langle v^2\rangle = \kappa\,
GM_{\rm init}/R$, from the scalar virial theorem.  Thus we find a
solution for $E\ge 0$ if $ M_\star \le M_{\rm init}/2$, i.e. as the
mass is reduced by 50\% or more ($\epsilon \le 1/2$), the remaining
system has zero or positive energy globally.

But if the law of averages applies, nothing stops a hard-bound core
forming at the expense of an expanding, loose envelope of stars, while
$E \ge 0$ for the system as a whole. We need to find out under which
circumstances this will occur. To do this we introduce an iterative
procedure based on the stellar distribution function in order to
determine the end-product fraction of bound stars.

\section{Distribution function: iterative scheme} 
As a simplifying assumption we take the sfe to be independent of
position in the cluster. Hence $\epsilon = $ constant (see Adams 2000 for 
a different approach). Thus gas and
stars are initially mixed in the same proportion throughout, however
note that $\epsilon$ does not fix the profiling of stellar masses with
radius, rather the total mass of gas made into stars~:

\[ \frac{\sum_{\rm i} ({\rm 
number\ of\ stars\ in\ mass\ bin\ } i) \times ({\rm stellar\ mass\ }\
m_i)}{m_{\rm gas}} = {\rm constant } \] 
at any radius $r$. Both gas and stars are distributed according to the
same distribution function which for a spherical system depends only
on energy, $f(E)$. The total system mass is then

\[ M = M_\star + M_{\rm gas} = \epsilon^{-1}\, M_\star = \int_{E_c}^{0}
f(E) {\rm d} E, \]
where $|E_c|$ is the maximum binding energy and
with a suitable normalisation of $f(E)$. The cluster potential
$\phi(r)$ may be written

\begin{equation} 
\phi(r ; M, R ) = \int_\infty^{r}\frac{{\rm d}u}{u^2} \int_0^u {4\pi
G\rho(w)w^2{\rm d}w} \equiv y(r/R)\, \frac{G M}{R}, \label{eq:phi}
\end{equation}
where $y(x)$ is a dimensionless function. Keeping $R$ constant while 
removing the gas instantly so the potential includes stars only, we have 

\begin{equation}
\phi(r; M, R ) \rightarrow \phi_\star = \epsilon\, \phi, \label{eq:phistar}
\end{equation}
and hence the fraction of stars at radius $r$ which now have positive
energy is

\begin{equation} 
1 - f_e = \frac{\displaystyle \int_{v_{e,\star}}^{v_e} f(v, r/R) \, v^2\,{\rm d}v }{ \displaystyle \int_{0}^{v_e} f(v, r/R)\, v^2\, {\rm d}v},
\label{eq:escape} \end{equation}
with $v_{e}$ the (local) escape velocity computed for $\phi$ (before
gas-expulsion), and similarly for $v_{e,\star}$ obtained for
$\phi_\star$.

Thus for a given d.f. and $\epsilon$, we may compute the quantity $1 -
f_e$ at all radii. Note that should the sfe be a function of
radius, the simple renormalisation (2) leading to $\phi_\star$ would
not apply, however (\ref{eq:escape}) may still be computed if the
stars' potential is given and $v_e$ known from the initial gas + stars
mixture.  The key step is to adjust the potential $\phi_\star$ itself,
since potential and velocity field do not match any longer. This is
normally computed using N-body integration to take into account the
full star-star interactions, and the redistribution of energy between
them in the time-dependent potential. If we thought that stars acquire
only little energy during the time that they escape, then the fraction
of stars remaining might be computed as follows.

Since the fraction (\ref{eq:escape}) of positive-energy stars is known
at all $r$, we recompute the gravitational potential counting only
stars with $E \le 0$ at each radius. Neglecting dynamical evolution, 
the cluster
radius $R$ is unchanged and hence the potential can be recomputed by
integration, from $\infty$, inwards. Once the new potential is known, a
fraction of the remaining stars will again be unbound by virtue of the
stars lost during the previous iteration. 
We  therefore repeat the procedure until
finally the cluster mass converges to a finite quantity, in which case
no more stars escape and the original distribution function is
depleted from all escapers in a self-consistent manner.

We consider three cases in detail, then turn briefly to numerical
calculations.

\subsection{Power-law d.f.}
The case $f(v) \propto v^{\beta - 2}$, where $\beta$ is constant,
provides a useful illustrative starting point. The d.f. is
truncated at the local escape velocity.  Then the velocity dispersion
$\sigma^2 \propto v_e^2 = 2\phi(r)$ at each radius. We find, on
substituting $\rho(w) \rightarrow \epsilon \rho(w)$, the new potential
given by (\ref{eq:phistar}) and $v_{e,\star}(r) = \sqrt{2
\phi_\star(r)} = \epsilon^{1/2} v_e(r)$. Equation (\ref{eq:escape}) 
 becomes (with $u \equiv v/v_e$)

\begin{equation} 
1 - f_e = \frac{\displaystyle \int_{v_{e,\star}}^{v_e} f(v, r/R) \,
v^2\,{\rm d}v }{ \displaystyle \int_{0}^{v_e} f(v, r/R)\, v^2\, {\rm
d}v} = \frac{\displaystyle \int_{\epsilon^{1/2}}^{1} f(u) \, u^2\,{\rm
d}u }{ \displaystyle \int_{0}^{1} f(u)\, u^2\, {\rm d}u} = 1 -
\epsilon^{(\beta+1)/2},
\label{eq:escapepower} 
\end{equation}
where we took $\beta > -1$. Thus $f_e = \epsilon^{(\beta+1)/2}$ is 
independent of radius. Repeating the procedure to take account of the
positive-energy stars, we substitute $\epsilon \rightarrow f_e \cdot
\epsilon$, etc, so that after $n$ iterations the net mass of bound
stars, $M^b_\star$, becomes

\begin{equation} M^b_\star = \epsilon^{[(\beta+1)/2]^n}\cdot \epsilon^{[(\beta+1)/2]^{n-1}} ... \epsilon^{(\beta+1)/2} \, \epsilon M \equiv 
\Pi_{k=1}^n \left( \epsilon^{[(\beta+1)/2]^k} \right) \, M_\star,
\label{eq:powerlaw}
\end{equation} 
(and similarly for the stellar density $\rho_\star[r]$ wrt $\rho[r]$).
The multiplicative operator $\Pi$ in (\ref{eq:powerlaw}) leads to a
non-zero (positive) value as $n\rightarrow\infty$ only when $\beta <
1$, since $\epsilon \le 1$. All d.f.'s with $\beta > 1$ lead to cluster
disruption, because the high-velocity range of the d.f. is too densely
populated, leading to catastrophic stellar loss after the expulsion of
any amount of gas.

\subsection{Plummer model}
We wish to compare our basic result (\ref{eq:powerlaw}) to a standard
fit to globular clusters. We consider the Plummer model, where $f(E)
\propto (-E)^{7/2}$. For this case the total system mass is finite but
infinite in extent; the velocity dispersion maximises at the centre,
as the density.  The velocity d.f., $f(v) \propto \ ( 1 - (v/v_e)^2
)^{7/2}$ (see Spitzer 1987); writing $p(\epsilon) = 105 -1210 \epsilon
+ 2104 \epsilon^2 - 1488 \epsilon^3 + 384 \epsilon^4$, we find

\begin{equation} 
1 - f_e = \frac{\displaystyle \int_{\epsilon^{1/2}}^{1} ( 1 - u^2
)^{7/2} \, u^2\,{\rm d}u }{ \displaystyle \int_{0}^{1}( 1 - u^2
)^{7/2} \, u^2\, {\rm d}u} = 1 +
\frac{\epsilon^{1/2}\sqrt{1-\epsilon}\, p(\epsilon) - 105 \sin^{-1}
\epsilon^{1/2} }{\frac{105\pi}{2} }. \label{eq:plummer}
\end{equation} 
We may repeat the procedure until $1 - f_e \rightarrow 0$ and no
additional stars are lost. We were not able to express the resulting
expression in simple form, however we note that, as in the first case,
the solution is independent of radius, hence the density profile is
simply renormalised at each radius. Therefore the potential
$\phi_\star$ and  escape velocity follow from
(\ref{eq:phistar}).

\subsection{Hernquist profile} 
To contrast with the smooth-density Plummer solution, we consider a
peaked Hernquist (1990) profile. The density $\rho$ and
one-dimensional velocity dispersion $\sigma$ vary radially according
to

\[ \rho(r\ {\rm or}\ x) \equiv \frac{M}{2\pi} \, \frac{1}{r} \, 
\frac{1}{(r+r_c)^3} = \frac{M}{2\pi r_c^3}\, \frac{1}{x}\,\frac{1}{(x+1)^3} =
\frac{1}{2\pi G r_c^2}\, \frac{\phi(x)}{x(x+1)^2}\ , \]
\[ \sigma^2(x) = \phi(x) \, x\, (1+x)^4\, \left\{ \ln \frac{1+x}{x} - 
\frac{1/4}{(1+x)^4} - \frac{1/3}{(1+x)^3} - \frac{1/2}{(1+x)^2} -
\frac{1}{1+x}\right\}, \]
with $r_c$ a free length fixing the point of the power-law turnover, and
$x \equiv r/r_c$. The velocity d.f., $f(v)$, is only constrained locally
by $\sigma(r)$; we thus have the freedom to choose any profile satisfying
$\sigma(r)$. We set $f(v) \propto v^2 \exp(-v^2/2\sigma^2) $, a
Maxwellian profile. This is found to give stable equilibria in N-body
calculations (Hernquist 1993).  Inserting this in (\ref{eq:escape})
yields

\begin{equation} 1 - f_e(\Psi_\star) = 1 - 
 \frac{ \sqrt{\Psi_\star}\, e^{-\Psi_\star} - \erf(\sqrt{2\Psi_\star})
   }{\sqrt{\Psi} e^{-\Psi} - \erf(\sqrt{2\Psi}) }, 
\label{eq:hernquist}
\end{equation} 
where the dimensionless potential $\Psi\equiv \phi/\sigma^2$ and
$\erf(x)$ is the error function. Note that although the dispersion
$\sigma^2 \propto \phi$ as before, here the fraction of
positive-energy stars depends on the local potential, and hence it is a
function of radius.  To compute the net fraction of bound star we must
therefore recompute the potential numerically for each evaluation of
$f_e$ in (\ref{eq:hernquist}).  This poses no problem since
$\phi(r\rightarrow\infty) \rightarrow 0$, and the density is known at
each step (though it is no longer a Hernquist profile).

\section{Results for three illustrative cases} 
Our results are shown in Fig. 1 for the three cases discussed
above. The solid lines show the solutions for the power-law d.f. with
$\beta = -3/4$ and 0. Note that in either case the fraction
$M^b_\star/M_\star$ of stars that remain bound is not dropping to zero until
$\epsilon$ itself is zero. The case $\beta = 0 $ corresponds to a flat
distribution, $f(v)v^2 {\rm d}v =$ constant $\times {\rm d}v$. 
Assuming only that no
star has velocity greater than the local escape velocity, the solution
(\ref{eq:powerlaw}) decreases only marginally faster than $\epsilon$
as $\epsilon \rightarrow 0$. 
By contrast, a Plummer or  Maxwellian-Hernquist model shows
 a sudden drop to a null fraction of bound stars for finite sfe : for
Plummer models we obtain a critical value $\epsilon \approx 0.445$,
while for the Hernquist model the fraction of bound stars 
exceeds 5\% or so until $\epsilon \approx 0.28$. In both cases the
iterative scheme converged to machine accuracy. 

Note that for the two self-consistent d.f.'s discussed here, the
power-law solutions provide an illustrative description of the
survival rate as $\epsilon$ approaches a critical value. The more
robust Hernquist model favours low-velocity stars (near the centre $\sigma \rightarrow 0$) and hence is better
fitted with the solution $\beta= -3/4$ in the range $0.5< \epsilon \le
1$. Figure 2 shows for this case that the fraction of bound stars 
$f_e$ remains larger near the core (see Fig. 2b). As a result, the 
initial density profile becomes steeper with radius. Since we have 
only made a selection by energy, the expectation is that the bound 
system indeed should be more peaked than initially. 

\setlength{\unitlength}{1in} 
\begin{picture}(4.1,4.1)(0,0) 
        \put(-.35,.1){\epsfxsize=0.7\textwidth\epsfysize=0.5\textheight
\epsfbox{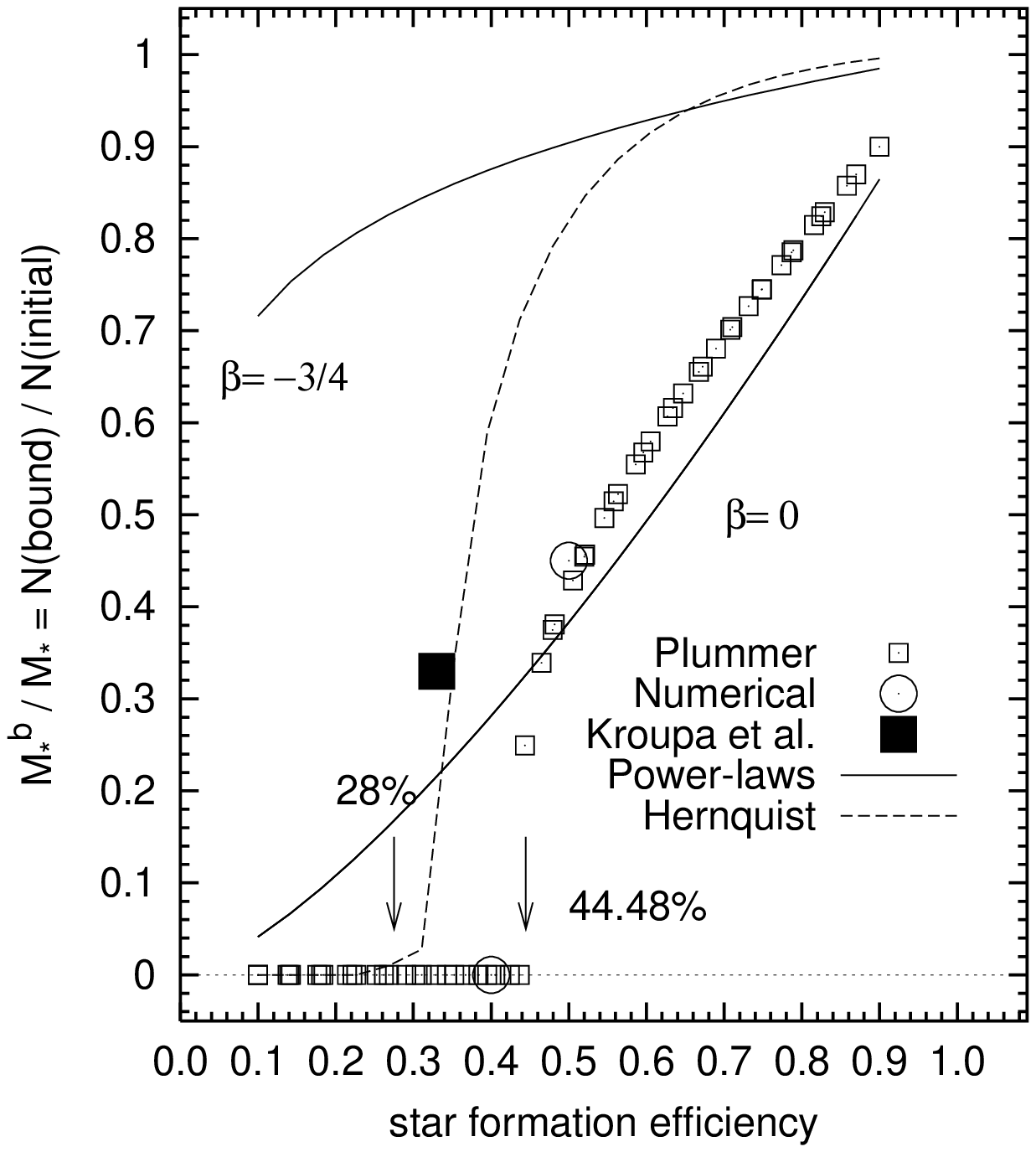}
                  }
       \put(3.3,3.0){
\begin{minipage}[t]{0.3\textwidth}{Figure 1. Ratio $M^b_\star/M_\star$
 of  bound 
 to initial stars as function of  star formation efficiency for the models of
Section~3.  }
\end{minipage} }
\end{picture}  

So far we only considered a mapping of static configurations under the
selection of particles according to ($\ref{eq:escape}$). In reality
stars must orbit in space before leaving the system, and they may
exchange gravitating energy with the background system in the process. 
N-body calculations are of some help here. N-body studies with
time-dependent potential (Lada, Margulis \& Deardorn 1984) or a star-gas mixture
(Geyer \& Burkert 2000) also find a critical sfe below which clusters
dissolve. We decided to conduct our own N-body calculations with a
collisionless grid code (Fellhauer et al. 2000) and 100,000 particle
equilibrium Plummer models.  Our analytic approach gives a critical
value for survival of $\epsilon = 0.4448$. We therefore setup two
N-body calculations with sfe = 50\% and 40\%. The fraction of stars
which remain bound in each case brackets the results derived from
(\ref{eq:plummer}) (see Fig. 1, large open circles). Notably we find
no indication that any stars remain for the case where
$\epsilon=0.40$, in agreement with the findings by others (Lada,
Margulis \& Deardorn 1984; Geyer \& Burkert 2000) for similar setups. 

\setlength{\unitlength}{1in} 
\begin{picture}(4,4.5)(0,0) 
        \put(-.5,.85){\epsfxsize=0.55\textwidth\epsfysize=0.5\textheight
\epsfbox{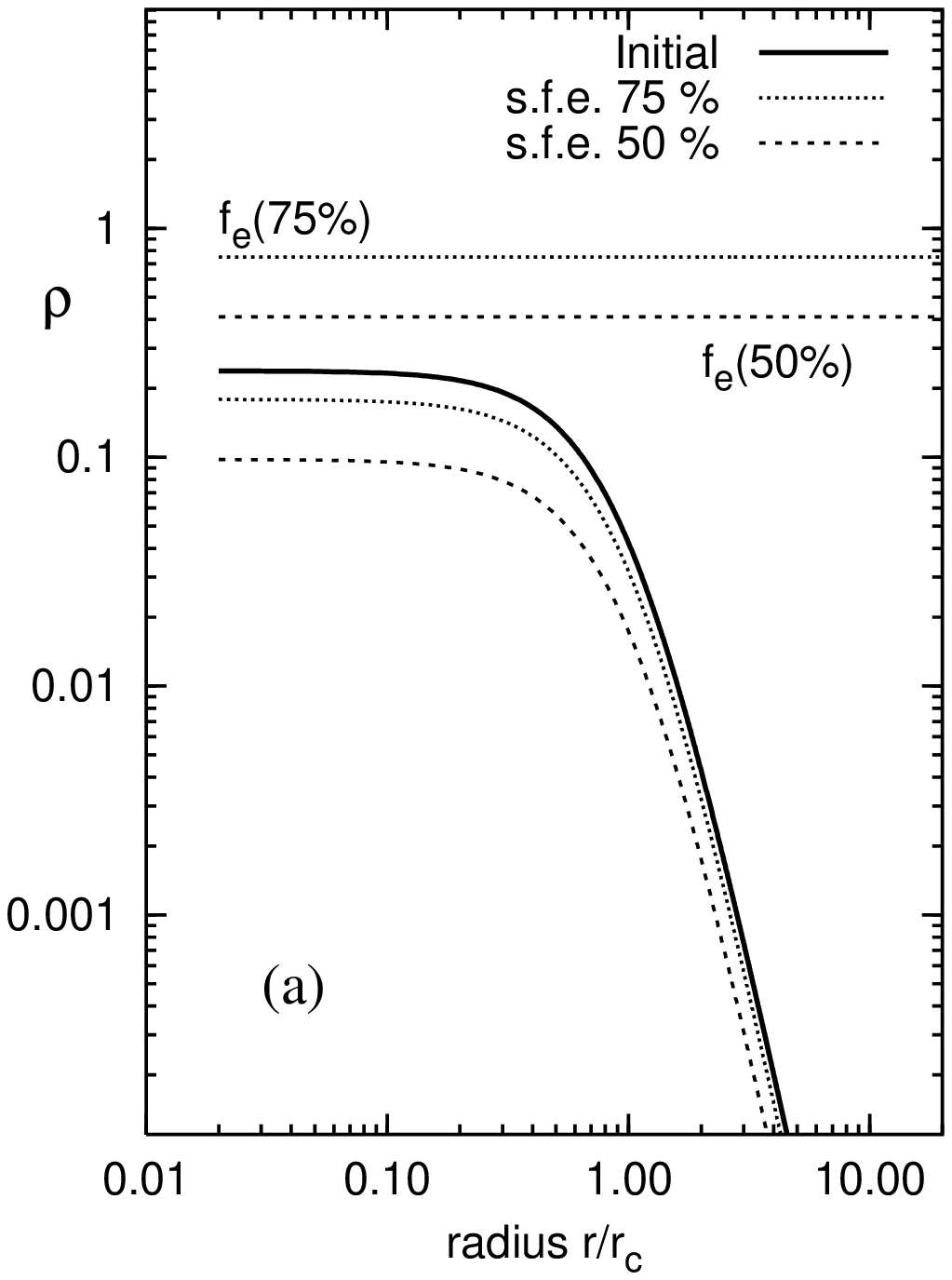}
                  }
        \put(2.25,.85){\epsfxsize=0.55\textwidth\epsfysize=0.5\textheight
\epsfbox{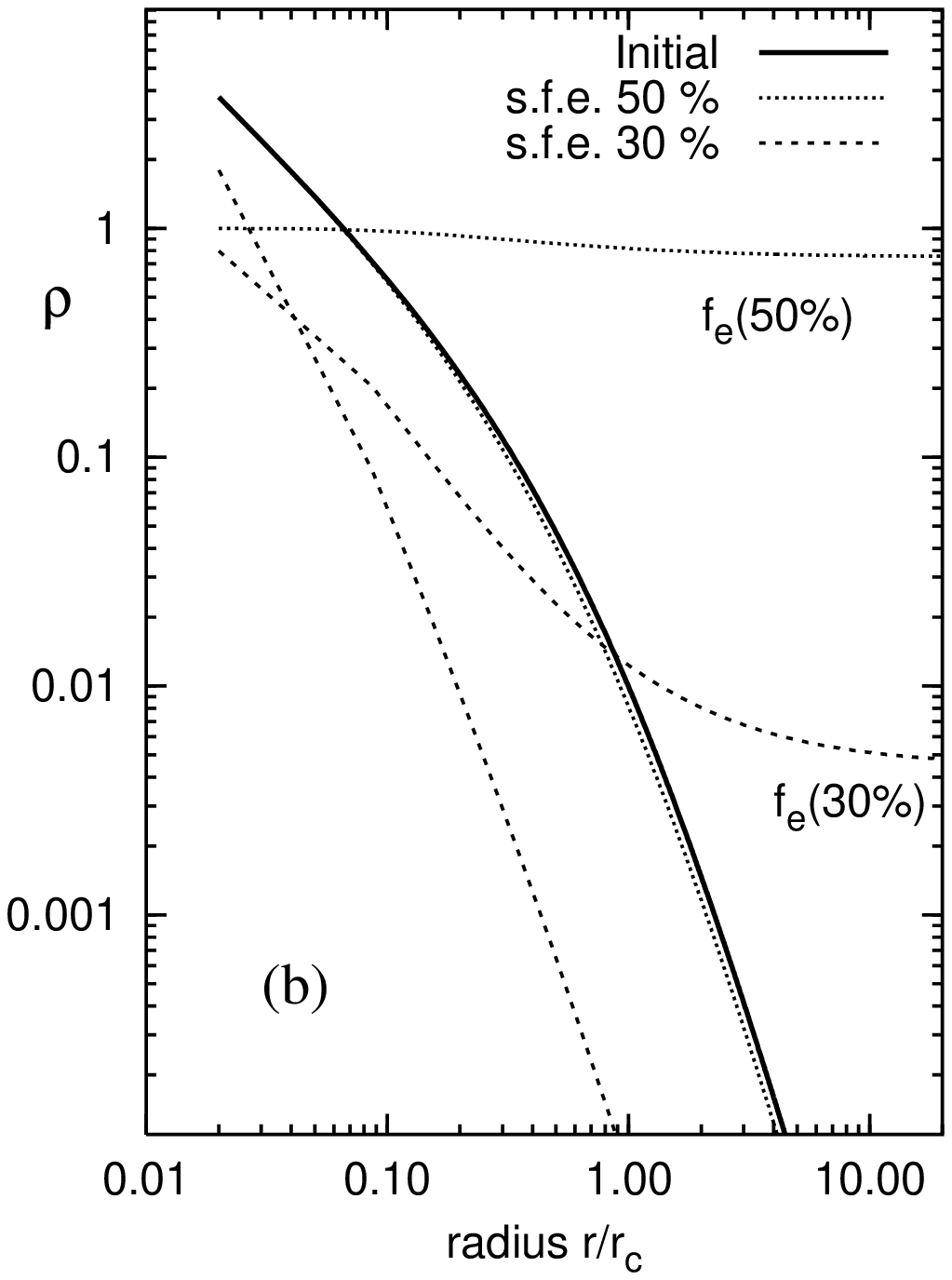} 
                  }
       \put(.0,0.55){
\begin{minipage}[t]{0.9\textwidth}
{Figure 2. Initial and final density profiles for (a) Plummer and (b)
Hernquist models for two values of the sfe $\epsilon$. The run of
$f_e$, i.e. the ratio of density to the initial density, is also shown
as a function of radius. Note how the Hernquist profile becomes
steeper for small $\epsilon$.  }  \end{minipage} }
\end{picture}  

\section{Time-evolution and other considerations}
To progress further, we note that only two timescales are important to
the problem, namely the cluster crossing time, $\tcr$, and the gas
removal timescale, $\tau$. We may understand the dynamics by comparing
$\tau$ to $\tcr$.  

\subsection{Collisional evolution, but short $\tau$} 
The key  lies with the
redistribution of kinetic energy between the stars. This can be
achieved on a short timescale by direct collisions or close
encounters, when the Safronov cross section $ d^2 \, \left( 1 +
\frac{1}{N^{2/3}} \right) $ is large (here $d^2$ is a star's geometric
cross section and $N$ the number of stars). This is especially true
for small-N open clusters with binaries and multiple stars, in which
case $d$ is set equal to their semi-major axis.  In this situation
collisional effects are never negligible and hence when  $\tau
\simeq \tcr $ or longer, the situation is not one of equilibrium, and
evolution must be tackled numerically. Kroupa, Aarseth \& Hurley
(2001) evolved an embedded 
 Plummer model with a high-precision direct-summation
$N$-body code and delayed, but near-to instantaneous, gas-removal. They 
find a fraction $\simeq 30\%$ of stars remain  despite
a low sfe of $\epsilon\approx0.3$. 
The results are similar to the analytical results 
for the Hernquist model (Fig.~1, black square), while we would have expected 
complete dissolution for collisionless evolution of a Plummer model. 
We are lead to
the conclusion that it is the compact core that develops as a
result of two-body relaxation {\it during the embedded phase} before
gas expulsion that leads to more robust clusters (see also Section~1).

\subsection{Collisionless evolution, but long $\tau$} 
Large-N systems possess a long two-body relaxation time $t_{\rm col} \simeq 
 \frac{N}{10\ln \gamma N} \tcr $. When $ \tcr \ll \tau \ll t_{\rm col}$, 
two-body effects may be neglected.
 In this case the age of the cluster may not have
allowed for global two-body relaxation, yet significant mass removal
will have occurred over several stellar orbits concentrated around the
centre and these orbits evolve adiabatically. Since adiabatic evolution 
 is a re-mapping of an orbit to itself, no stars on such orbits are lost. 
  Thus for finite or large $\tau$, we anticipate the survival rate of 
clusters to be intimately linked to their properties at birth, such as 
 what 
family of orbits are present initially. 
\newline 

 The future of this programme will see
additional high-accuracy $N$-body computations being performed to
address how relatively important the three hypothesis raised in
Section~1 are for the formation of bound clusters. Specifically, the
formation of sub-condensations as a result of encounters during the
radially expanding flow will be addressed in detail.

\end{document}